# A Brief Study of Computer Network Security Technologies

Tulasi Udupa A

Sushma Jayaram

Shreya Ganesh Hegde

Under the guidance of:

Dr. C. S. Jayasheela

Department of ISE

Bangalore Institute of Technology

## Abstract

The rapid development of computer network system brings both a great convenience and new security threats for users. Network security problem generally includes network system security and data security. Specifically, it refers to the reliability of network system, confidentiality, integrity and availability of data information in the system. This paper introduces the significance of network security systems and highlights related technologies, mainly authentication, data encryption, firewall and antivirus technology. Network security problems can be faced by any network user, therefore we must greatly prioritize network security, try to prevent hostile attacks and ensure the overall security of the network system.

## Introduction

The exponential growth in the popularity and adoption of the computer network, especially the emergence of the Internet, results in easy access to a vast variety information applications, widely spread across the world. However, information of all nature are transmitted and stored in the public communication network, which may be illegally wiretapped, intercepted, tampered or damaged by attackers with a variety of the purposes, thus resulting in an immeasurable loss. The threats on network security mainly display in: illegal access, pretending to be legal users, destroying data, listening in line and using network to transport virus, etc. With the network security problem growing increasingly prominent, whether the network security problem can be solved has become one of the key factors restricting the development of network. Network security problems generally include network system security and data security. Network system security is to prevent the system from illegal attack, access and destruction; while data security is mainly to prevent confidential and sensitive data from being steal or illegal copied. The computer network security problem is related to many fields, such as computer technology, communication technology, mathematics, cryptography, information theory, management and low. From the perspective of different fields, there are different solutions for network security problem, and these solutions should be integrated to solve the problem. This paper mainly introduces network security background and its necessity in Section 2, discusses network security technologies in Sections 3 through 6 and concludes the paper with Section 7.

## Background and Necessity of Network Security

Network security refers to the reliability of a network system as well as the confidentiality, integrity and availability of information in the system.

ITU-TX.800 standard defines the network security logically from three aspects:

1. Security Attack: acts that damage information in the network, including information denial service, message change, message replay, camouflage, traffic analysis, message release, etc.

2. Security Mechanism: mechanisms designed for detecting and preventing security attacks and for system recovery. It includes encryption, digital signature, access control, data integrity, authentication exchange, traffic padding, routing control, notarization and other mechanisms.

3. Security Service: refers to the services to defend against the security attacks and improve the data process and information transmission security with one or more security mechanisms. It includes the peer entity authentication, data source authentication, access control, confidentiality, traffic flow confidentiality, data integrity, non-repudiation and availability, etc.

Network security and associated problems come into play at all layers of the network, as discussed below.

- At the physical layer, it is mainly to prevent the damage, eavesdropping and attack on the physical path.
- At the data link layer, it ensures that the data transferred through the network link is protected from eavesdropping with techniques such as VLAN in LAN and encryption communication in WAN.
- Network layer security ensures that the network provides authorization services only for authorized users, to guarantee correct network routing and avoid eavesdropping or being blocked.

- Transport layer security ensures the security of information flow.
- Operating System security refers to the security of operating system access control, such as database server, mail server, and Web server. Due to the complexity of operating systems, multiple technologies are always applied to enhance the operating system security.
- Application System security is also of great importance as the ultimate purpose of network system is to serve the users. It refers to the security services provided by the application platform, such as communication content security, including two-way communication authentication and the auditing system.

Network security is necessary to uphold confidentiality, authenticity, integrity, dependability, availability and auditability, which are further outlined below.

- Confidentiality: the system should only provide information to the authorised users.
- Authenticity: it should be guaranteed for the receiver that the information is from the claimed source.
- Integrity: the system should only allow the authorised users to modify the information, thus ensuring that the information is complete.
- Dependability: it should prevent the sender or receiver from denying the transmitted or received messages.
- Availability: authorised users should get the required information resource services from the system.
- Auditability: all the activities in the system related to security should be capable of being reviewed.

## Authentication

Authentication is the process of identifying users that request access to a system, network, or device. In the era of mobile networks and ubiquitous computing, traditional threat distinctions fail to address the growing risk of internal malware infiltration. Publicly accessible networks, like those found in libraries and coffee shops, expose secure networks to potential threats from unsecured devices brought in by users. These devices often lack updated software, bypassing conventional external firewall defenses. Mobile clients pose a dual risk: susceptibility to infection and the potential to strain network resources by propagating malware. Even if networks are monitored, mobile clients can connect briefly enough to spread malware before detection. Personal firewalls offer limited protection in this diverse environment. Authentication mechanisms like LEAP, though used for access control, fall short in preventing internal malware spread. Users may authenticate correctly but inadvertently bring infected devices from other networks. Most networks lack internal security measures to prevent such compromises, and resource-intensive intrusion detection systems often don't monitor local communication.

A malware defense system for authentication proposed in [5] focuses on isolating vulnerable and infected machines preemptively. Comprising three main components as shown in Fig. 1 [5]—vulnerability detection, quarantine enforcement, and policy management—the system aims to safeguard networks effectively. Security authentication complements user authentication to assess system risks accurately. Upon connection, machines undergo vulnerability scans to determine appropriate network access levels, adjusted based on perceived threats. Quarantine measures are tailored to restrict malicious activity while allowing essential functions, facilitating controlled operation until fixes are implemented. Integration with standard networking technologies enables effective network and MAC layer isolation, crucial for containment. The policy manager oversees continuous scanning, assessment, and quarantine placement, ensuring machines are adequately secured and managed over time. By integrating security authentication and proactive quarantine measures, the system offers a comprehensive defense against evolving malware threats.

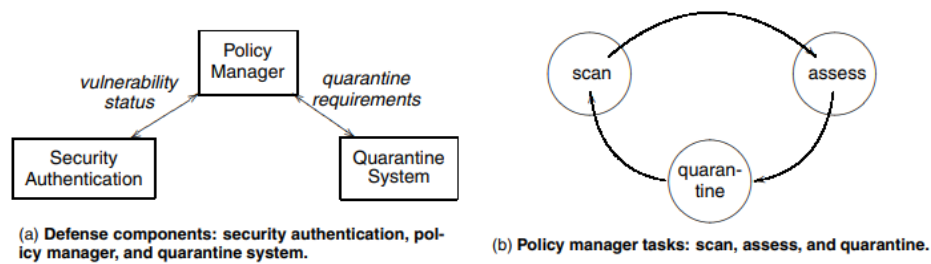

Figure 1. Malware defense system components and tasks

## Data Encryption

Encryption, an ancient practice, serves as the foundation of data security, employing codes and ciphers to safeguard messages from unauthorized access or interception. In modern data security, encryption addresses three critical concerns: privacy, authentication, and integrity. By encoding messages, encryption ensures that only authorized parties can access and decipher the information, while also verifying the identities of the sender and receiver, and guaranteeing that the content remains unaltered during transmission.

SSL (Secure Sockets Layer) stands as a standardized technology facilitating secure communication between web servers and Internet browsers, or between mail servers and mail clients such as Outlook. This protocol establishes a secure connection, preserving the confidentiality and integrity of data exchanged between the server and the user's browser. It employs cryptographic algorithms to encrypt plaintext transmitted over insecure channels, ensuring data confidentiality throughout the communication process. The connection between the web browser and webserver using the SSL system is shown in Fig. 2 [2].

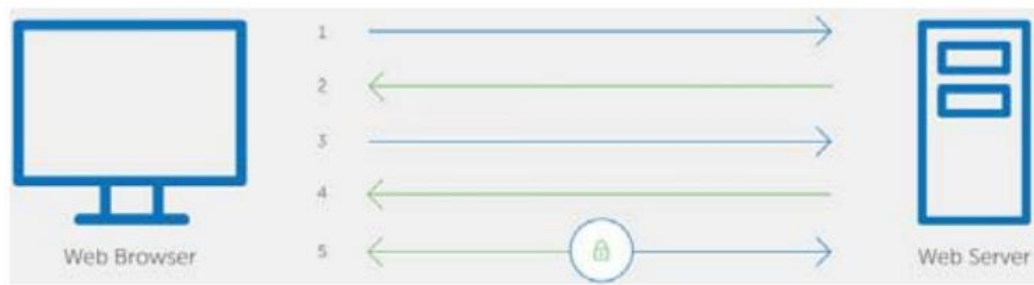

Figure 2. The connection between the web browser and webserver using

SSL operates through a process known as the SSL handshake, wherein clients and servers authenticate each other using digital certificates and exchange cryptographic keys for encrypting and decrypting data. Digital certificates issued by Certification Authorities (CAs) validate the identities of the parties involved in the communication and confirm the authenticity of the server. SSL employs a combination of symmetric and asymmetric encryption techniques, leveraging the speed of symmetric encryption and the robust authentication capabilities of public key encryption. The SSL handshake, executed invisibly to users, generates secure encryption keys, ensuring that data encrypted with the public key can only be decrypted with the corresponding private key, and vice versa, thereby safeguarding the confidentiality and integrity of transmitted data [2].

**Firewall**

Firewall serves as a fundamental infrastructure for providing information security services and ensuring network and information security. Positioned as a barrier between external and internal networks, as illustrated in Fig. 3 [3], firewalls operate as a combination of software and hardware, effectively segregating information flow between internal and external networks, safeguarding internal network information security and protecting user information within the network. Broadly categorized, firewalls manifest in two systems: Packet filtering firewall (Packet Filter) and proxy firewall (Application Gateway). Packet filtering technology operates at the network layer, where defined filtering rules are applied to inspect each packet passing through. If a packet aligns with the filtering rules, it is permitted to traverse; otherwise, the data packet is rejected. Conversely, proxy firewalls, employing a proxy server positioned between clients and Internet servers, offer complete isolation between internal and external networks. With distinct functionalities for various client applications such as FTP, HTTP, Telnet, and SMTP, firewalls primarily govern and manage external network access to the internal network.

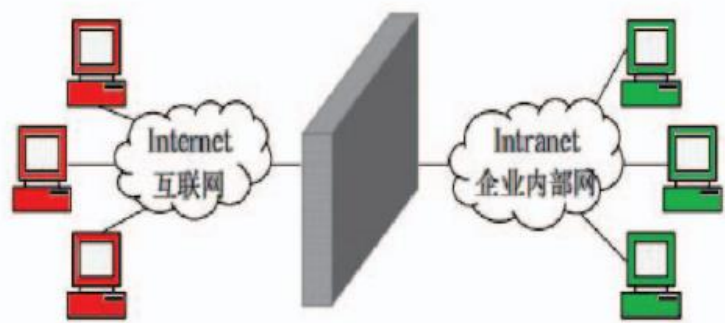

Figure 3. Firewall technology

There are several advantages of employing firewalls: they efficiently and conveniently protect internal networks, as all inbound and outbound information must pass through firewalls, serving as the sole access point. In the event of firewall intrusion or malfunction, it can forcibly sever the connection between internal and external networks, ensuring internal network safety, while also issuing early warnings and tracing the origin of suspicious activities to enable prompt network management responses.

Firewall technologies, however, also exhibit notable shortcomings in contemporary network security in the era of big data [3]. While firewalls serve as foundational safeguards against external threats, they remain peripheral in their monitoring capabilities, primarily focusing on user operations and network layer access control. Consequently, firewalls struggle to ensure real-time monitoring of information, leaving communication content vulnerable and increasing network interference.

**Antivirus**

Virus is a computer program that replicates itself and corrupts the system data it is loaded onto. These viruses replicate themselves and then infect other folders. A virus can quickly use up all the memory by replicating itself, or can transmit across networks and bypass security systems, compromising the security of all the systems in the network.

An antivirus is a computer program/interpretation system that can be used to scan files to detect and eliminate computer virus. This system is operating system specific. A basic structure of the system is outlined in Fig. 4 [4].

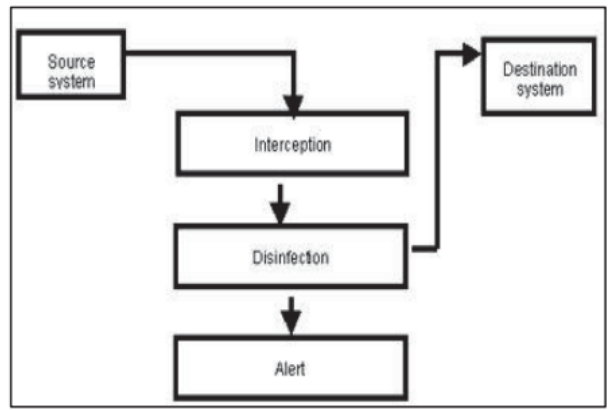

Figure 4. Basic structure of antivirus

There are many techniques for detecting viruses [4]. **Signature based detection** uses an examined file to create a static fingerprint of known malware in the form of stream of bytes or cryptographic hash. A limitation is that this method is unable to flag malicious files for which signatures have not been developed, allowing attackers to frequently mutate their creations by changing the file's signature. **Heuristic based detection** detects new malware by statically examining files for suspicious characteristics with the help of an algorithm. It can look for the presence of rare instructions in the examined file. However, it can flag legitimate files as malicious too. **Behavioral detection** monitors how the program executes, rather than merely emulating its execution. **Cloud-based detection** identifies malware by collecting data from clients while analyzing it on the provider's cloud, instead of performing the analysis locally. The vendor's cloud engine can derive patterns related to malware characteristics by correlating data from multiple systems. **SandBox** detection isolates and creates an emulated environment for programs, and analyzes its behavior. If the behavior is destructive, the user is warned before running the program.

Types of viruses [4] include the **File Infector/Directory Viruses** which infects program files by changing their paths that indicate the location of the executable code, like .EXE and .COM files. **Boot Sector Viruses** attach themselves to the boot record of disks and attacks when the user attempts to start up from the infected disk. The **Master boot record viruses** similarly infect the Master Boot Record. **Multi-Partite Viruses** shares some characteristics of boot sector viruses and file viruses and can infect .COM and .EXE files, and the boot sector of the computer's hard drive. Distributed through infected media, they usually hide in the memory and are very difficult to repair. **Macro Viruses** infect data files created by application programs that contain macros, like Microsoft Excel. **Trojan** viruses are malicious, security-breaking program, which causes damage silently. **Worms** are a special type of virus that has the ability to self-replicate and use memory area, but cannot attach it to other programs. The **Encrypted Viruses** consist of encrypted, malicious code which makes it difficult for antivirus to detect them.

## Conclusion

In conclusion, network security is a critical aspect of preserving the integrity, confidentiality, and availability of data in today's interconnected environment. Our examination of authentication, data encryption, firewalls, and antivirus technologies underscores their significant roles in bolstering network defenses. Despite their importance, each technology has inherent limitations, such as vulnerabilities in authentication processes, encryption weaknesses, monitoring challenges with firewalls, and antivirus software lagging behind emerging threats. However, these challenges present opportunities for improvement. Advancements in machine learning and artificial intelligence can enhance authentication processes, while better encryption protocols and key management strategies can bolster data protection. Next-generation firewalls can integrate more robust intrusion detection and prevention features, and leveraging cloud-based threat intelligence can empower antivirus solutions to swiftly identify and mitigate malware threats. A comprehensive approach that combines robust technologies with proactive threat intelligence, rigorous policies, and ongoing user education is essential for mitigating evolving cyber threats and ensuring a secure digital environment for all stakeholders.

## References

1. F. Yan, Y. Jian-Wen and C. Lin, "Computer Network Security and Technology Research," 2015 Seventh International Conference on Measuring Technology and Mechatronics Automation, Nanchang, China, 2015, pp. 293-296, doi: 10.1109/ICMTMA.2015.77. Available: https://ieeexplore.ieee.org/document/7263569
2. Roza Dastres, Mohsen Soori. Secure Socket Layer (SSL) in the Network and Web Security. International Journal of Computer and Information Engineering, In press, 14 (10), pp.330-333. ffhal-03024764f. Available: https://hal.science/hal-03024764
3. Q. Zhang, Z. Luo and J. Zhang, "Analysis of the Application of Firewall and Intrusion Detection Technology in Network Security in the Era of Big Data," 2023 2nd International Joint Conference on Information and Communication Engineering (JCICE), Chengdu, China, 2023, pp. 167-171, doi: 10.1109/JCICE59059.2023.00042. Available: https://ieeexplore.ieee.org/document/10217276
4. Kaur, G. (2016). Network security: Anti-virus. International Journal of Advanced Research in Computer Science, 7(6). Available: https://www-proquest-com-vtuconsortia.knimbus.com/scholarly-journals/network-security-anti-virus/docview/1875134290/se-2
5. J. V. Antrosiom and E. W. Fulp, "Malware defense using network security authentication," Third IEEE International Workshop on Information Assurance (IWIA'05), College Park, MD, USA, 2005, pp. 43-54, doi: 10.1109/IWIA.2005.11. Available: https://ieeexplore.ieee.org/document/1410701